\documentclass[11pt]{article}
\usepackage{moriond,epsfig}
\usepackage{w-greek}
\usepackage{amsmath,amsthm,amscd,amssymb}
\usepackage{latexsym,wasysym}
\usepackage{graphicx,epsfig}

\def\Journal#1#2#3#4{{#1} {\bf #2}, #3 (#4)}

\def\PRD{{\em Phys. Rev.} D}

\def\rfr#1{Eq.~\ref{#1}}

\def\eqi{\begin{equation}}
\def\eqf{\end{equation}}
\def\eqia{\begin{eqnarray}}
\def\eqfa{\end{eqnarray}}
\def\rp#1#2{{#1\over#2}} \def\lb#1{\label{#1}}

\begin{document}
\vspace*{4cm}
\title{ON THE ANOMALOUS INCREASE OF THE LUNAR ECCENTRICITY}

\author{ L. IORIO }

\address{Ministero dell'Istruzione, dell'Universit\`{a} e della Ricerca (M.I.U.R.)\\ Viale Unit\`{a} di Italia 68, 70125, Bari (BA), Italy.}

\maketitle\abstracts{
Possible explanations of the recently reported anomalous increase of the eccentricity of the lunar orbit are sought in terms of classical Newtonian mechanics, general relativity, and long-range modifications of gravity.}

%
%
Anderson and Nieto, { in a  recent review\cite{And010} of some astrometric anomalies  detected in the solar system by some independent groups, mentioned also} an anomalous secular increase of the eccentricity $e$ of the orbit of the Moon
\eqi \dot e_{\rm meas} = (9\pm 3)\times 10^{-12}\ {\rm yr^{-1}}\lb{ecce_meas}\eqf based on an analysis of a long LLR data record spanning 38.7 yr  performed by Williams and Boggs\cite{WiBo} with the dynamical force models of the DE421 ephemerides\cite{de421a,de421b} including all the known relevant Newtonian and Einsteinian effects. Notice that \rfr{ecce_meas} is statistically significant at a $3\sigma-$level.
{The first account\cite{JGR} of this effect appeared in 2001 by Williams {\em et al.}, who gave an extensive discussion of the state-of-the-art in modeling the tidal dissipation in both the Earth and the Moon. Later, Williams and Dickey\cite{hjk}, relying upon the  2001 study\cite{JGR}, released an anomalous eccentricity rate as large as $\dot e_{\rm meas}=(1.6\pm 0.5)\times 10^{-11}$ yr$^{-1}$. }
Anderson and Nieto\cite{And010} commented that \rfr{ecce_meas} is not compatible with present, standard knowledge of the dissipative processes in
the interiors of both the Earth and Moon, which were, actually, modeled by Williams and Boggs\cite{WiBo}.
%
%

Naive, dimensional evaluations of the effect caused on $e$ by an additional anomalous acceleration $A$ can be made by noticing that
\eqi \dot e\simeq \rp{A}{na},\lb{misl}\eqf with
\eqi na = 1.0\times 10^3\ {\rm m\ s^{-1}}=3.2\times 10^{10}\ {\rm m\ yr^{-1}}\eqf
for the geocentric orbit of the Moon, whose mass is denoted as $m$. In it, $a$ is the orbital semimajor axis, while $n\doteq\sqrt{\mu/a^3}$ is the Keplerian mean motion in which $\mu\doteq GM(1+m/M)$ is the gravitational parameter of the Earth-Moon system: $G$ is the Newtonian constant of gravitation and $M$ is the mass of the Earth. It turns out that an extra-acceleration  as large as  \eqi A\simeq 3\times 10^{-16}\ {\rm m\ s}^{-2}=0.3\ {\rm m\ yr^{-2}}\lb{mutuu}\eqf would satisfy \rfr{ecce_meas}. In fact, a mere order-of-magnitude  analysis based on \rfr{misl} would be inadequate to infer meaningful conclusions:  finding simply that this or that dynamical effect induces an extra-acceleration of the right order of magnitude may be highly misleading. Indeed,  exact calculations of the secular variation of $e$ caused by such putative promising candidate extra-accelerations $A$ must be performed with standard perturbative techniques in order to check if they, actually, cause an
averaged non-zero change of the eccentricity. Moreover,  it may well happen, in principle, that the resulting analytical expression for $\left\langle\dot e\right\rangle$ retains multiplicative factors $1/e^j, j=1,2,3,...$ or $e^j, j=1,2,3...$ which would notably alter the size of the found non-zero secular change of the eccentricity  with respect to the expected values according to \rfr{misl}.
%

It is well known that a variety of theoretical paradigms\cite{Adel,Berto} allow for Yukawa-like deviations\cite{Bur}  from the usual Newtonian inverse-square law of gravitation. The Yukawa-type correction to the  Newtonian gravitational potential $U_{\rm N}=-\mu/r$, where $\mu\doteq GM$ is the gravitational parameter of the central body  which acts as source of the supposedly modified gravitational field,
 is \eqi  U_{\rm Y}=-\rp{\alpha\mu_{\infty}}{r}\exp\left(-\rp{r}{\uplambda}\right),\lb{uiu}\eqf in which $\mu_{\infty}$ is the gravitational parameter evaluated at distances $r$ much larger than the scale length $\uplambda$.
In order to compute the long-term effects of \rfr{uiu} on the eccentricity of a test particle it is convenient to adopt the Lagrange perturbative scheme\cite{BeFa}.
 In such a framework,  the equation for the long-term variation of $e$ is\cite{BeFa}
 \eqi \left\langle\rp{d e}{dt}\right\rangle = \rp{1}{na^2}\left(\rp{1-e^2}{e}\right)\left( \rp{1}{\sqrt{1-e^2}}\rp{\partial \mathcal{R}}{\partial\omega} -  \rp{\partial \mathcal{R}}{\partial\mathcal{M}} \right),\lb{eLag}\eqf where $\omega$ is the argument of pericenter,  $\mathcal{M}$ is the mean anomaly of the test particle, and $\mathcal{R}$ denotes the average of the perturbing potential over one orbital revolution.
 In the case of a Yukawa-type perturbation, \rfr{uiu} yields
 \eqi\left\langle  U_{\rm Y}\right\rangle=-\rp{\alpha \mu_{\infty}\exp\left(-\rp{a}{\uplambda}\right)}{a}I_0\left(\rp{ae}{\uplambda}\right), \lb{poti}\eqf
where $I_0(x)$ is the modified Bessel function of the first kind
$I_q(x)$ for $q=0$. An inspection of \rfr{eLag} and \rfr{poti}  immediately tells us that there is no secular variation of $e$ caused by an anomalous Yukawa-type perturbation.
%
%

The size of the general relativistic Lense-Thirring\cite{LT} acceleration experienced by the Moon because of the Earth's angular momentum\cite{IERS} $S=5.86\times 10^{33}$ kg m$^2$ s$^{-1}$  is
just
\eqi A_{\rm LT}\simeq \rp{2v G S}{c^2 a^3}=1.6\times 10^{-16}\ {\rm m\ s^{-2}} = 0.16\ {\rm m\ yr^{-2}},\eqf i.e. close to \rfr{mutuu}.
On the other hand, it is well known that the  Lense-Thirring effect does not cause long-term variations of the eccentricity. Indeed, the integrated shift of $e$ from an initial epoch corresponding to $f_0$ to a generic time corresponding to $f$ is\cite{Soff}
\eqi \Delta e = -\rp{2 G S\cos I{^{'}}\left(\cos f-\cos f_0\right)}{c^2 n a^3\sqrt{1-e^2}},\lb{letie}\eqf in which  $I^{'}$ is  the inclination of the Moon's orbit with respect to the Earth's equator and $f$ is the true anomaly. {From \rfr{letie} it straightforwardly follows that after one orbital revolution, i.e. for $f\rightarrow f_0+2\pi$, the long-term gravitomagnetic shift of $e$ vanishes.}
%
%

A promising candidate for explaining the anomalous increase of the lunar eccentricity
   is, at least in principle, a  trans-Plutonian  massive body X of planetary size located in the remote peripheries of the solar system. Indeed,  the perturbation induced by it would, actually, cause a non-vanishing long-term variation of $e$. Moreover, since it depends on the spatial position of X in the sky and on its tidal parameter
\eqi \mathcal{K}_{\rm X}\doteq \rp{Gm_{\rm X}}{d_{\rm X}^3},\eqf where $m_{\rm X}$ and $d_{\rm X}$ are the mass and the distance of X, respectively, it may happen that a suitable combination of them is able to reproduce  \rfr{ecce_meas}.
Let us recall that{, in general,} the perturbing potential {felt by a test particle orbiting a central body} due to a very distant, pointlike mass can be cast into the following quadrupolar form
\eqi U_{\rm X} = \rp{\mathcal{K}_{\rm X}}{2}\left[r^2 -3\left( \vec{r}\cdot{\hat{l}} \right)^2\right],\lb{ux}\eqf
where
 ${\hat{l}}=\left\{l_x,l_y,l_z\right\}$ is a unit vector directed towards X determining its position in the sky.
In \rfr{ux} $ \vec{r}=\left\{x,y,z\right\}$ {is the geocentric position vector of} the perturbed particle{, which, in the present case, is the Moon}.
Iorio\cite{Iorio} has recently shown that the average of \rfr{ux} over one orbital revolution {of the particle} is
\begin{equation}
\left\langle U_{\rm X} \right\rangle   =  \rp{{\mathcal{K}_{\rm X}}a^2}{32}\mathcal{U}\left(e,I,\Omega,\omega; {\hat{l}}\right),\lb{us}
\end{equation}
where  $\mathcal{U}\left(e,I,\Omega,\omega; {\hat{l}}\right)$ is a complicated function of its arguments\cite{Iorio}: $\Omega$ is the longitude of the ascending node and $I$ is the inclination of the lunar orbit to the ecliptic.
In the integration yielding \rfr{us} ${\hat{l}}$ was kept fixed over one orbital revolution of the {Moon}, as it is reasonable given the assumed large distance of X with respect to it. \rfr{eLag}, applied to \rfr{us}, straightforwardly yields
\eqi\left\langle\dot e\right\rangle  = \rp{15\mathcal{K}_{\rm X}e\sqrt{1-e^2}}{16n}\mathcal{E}\left(I,\Omega,\omega; {\hat{l}}\right).\lb{eccecazzo}\eqf
Also $\mathcal{E}\left(I,\Omega,\omega; {\hat{l}}\right)$ is an involved function of the orientation of the lunar orbit in space and of the position of X in the sky\cite{Iorio}.
Actually, the expectations concerning X are doomed to fade away. Indeed, apart from the modulation introduced by the presence of the time-varying $I,\omega$ and $\Omega$ in \rfr{eccecazzo}, the values for the tidal parameter which would allow to obtain \rfr{ecce_meas} are too large for all the conceivable positions $\left\{\beta_{\rm X},\lambda_{\rm X}\right\}$ of X in the sky. This can easily be checked by keeping $\omega$ and $\Omega$ fixed at their J2000.0 values as a first approximation. Indeed, Iorio\cite{Iorio} showed that the  physical and orbital features of X postulated by  two recent plausible theoretical scenarios\cite{Lyka,Mate} for X would induce long-term variations of the lunar eccentricity much smaller than \rfr{ecce_meas}.
Conversely, it turns out that a tidal parameter as large as
\eqi \mathcal{K}_{\rm X}=4.46\times 10^{-24}\ {\rm s^{-2}}\lb{cazzara}\eqf would yield the result of \rfr{ecce_meas}. Actually, \rfr{cazzara} is totally unacceptable since it corresponds to distances of X as absurdly small as $d_{\rm X}=30$ au for a terrestrial body, and $d_{\rm X}=200$ au for a Jovian mass.

An empirical explanation of \rfr{ecce_meas} can be found by assuming that, in addition to the usual Newtonian inverse-square law for the gravitational acceleration  imparted to a test particle by a central body orbited by it, there is also a small radial extra-acceleration of the form
\eqi A= k H_0 v_r.\lb{hubacc}\eqf In it $k$ is a positive numerical parameter of the order of unity to be determined from the observations, $H_0=(73.8\pm 2.4)$ km s$^{-1}$ Mpc$^{-1}=(7.47\pm 0.24)\times 10^{-11}$ yr$^{-1}$  is the Hubble parameter at the present epoch\cite{hubb}, defined in terms of the time-varying cosmological scaling factor $S(t)$ as $H_0\doteq \left.\dot S/S\right|_0$, and $v_r$ is the component of the velocity vector $\vec{v}$ of the test particle's proper motion about the central body along the common radial direction.
Indeed, a straightforward application of the Gauss perturbative equation for $e$ to \rfr{hubacc} yields
\eqi \left\langle \dot e \right\rangle  = kH_0 \rp{\left(1-e^2\right)\left(1-\sqrt{1-e^2}\right)}{e}.\lb{syste}\eqf
Since $e_{\rm Moon}=0.0647$,  \rfr{syste} can reproduce \rfr{ecce_meas} for $2.5\lesssim k\lesssim 5$.
Here we do not intend to speculate too much about possible viable physical mechanisms yielding the extra-acceleration of \rfr{hubacc}.
It might be argued  that, reasoning within a cosmological framework, the Hubble law may give \rfr{hubacc} for $k=1$ if the proper motion of the particle about the central mass is taken into account in addition to its purely cosmological recession which, instead, yields the well-known local extra-acceleration of tidal type $A_{\rm cosmol} = -q_0 H^2_0 r,$ where $q_0$ is the deceleration parameter at the present epoch.
%
\section*{Acknowledgments}
I gratefully acknowledge the financial support by the MORIOND scientific committee

\end{document}